\documentclass[english,twocolumn,english,aps, prl, reprint]{revtex4-1}
\usepackage[T1]{fontenc}
\usepackage[latin9]{inputenc}
\usepackage{amstext}
\usepackage{graphicx}

\makeatletter

\DeclareRobustCommand{\greektext}{%
  \fontencoding{LGR}\selectfont\def\encodingdefault{LGR}}
\DeclareRobustCommand{\textgreek}[1]{\leavevmode{\greektext #1}}
\DeclareFontEncoding{LGR}{}{}
\DeclareTextSymbol{\~}{LGR}{126}

\@ifundefined{textcolor}{}
{%
 \definecolor{BLACK}{gray}{0}
 \definecolor{WHITE}{gray}{1}
 \definecolor{RED}{rgb}{1,0,0}
 \definecolor{GREEN}{rgb}{0,1,0}
 \definecolor{BLUE}{rgb}{0,0,1}
 \definecolor{CYAN}{cmyk}{1,0,0,0}
 \definecolor{MAGENTA}{cmyk}{0,1,0,0}
 \definecolor{YELLOW}{cmyk}{0,0,1,0}
 }

\usepackage{graphicx}
\usepackage{amsmath}
\usepackage[colorlinks=true]{hyperref}

\makeatother

\usepackage{babel}

\begin{document}

\title{Evanescent Gain in {}``Trapped Rainbow'' Negative Refractive Index
Heterostructures}

\author{Edmund I. Kirby, Joachim M. Hamm, Tim Pickering, Kosmas L. Tsakmakidis,
and Ortwin Hess}

\email{o.hess@surrey.ac.uk}

\affiliation{Advanced Technology Institute and Department of Physics, Faculty
of Engineering and Physical Sciences, University of Surrey, Guildford,
GU2 7XH, United Kingdom}
\begin{abstract}
We theoretically and numerically analyze a five-layer {}``trapped
rainbow'' waveguide made of a passive negative refractive index (NRI)
core layer and gain strips in the cladding. Analytic transfer-matrix
calculations and full-wave time-domain simulations are deployed to
calculate, both in the frequency- and in the time-domain, the losses
or gain experienced by complex-wavevector and complex-frequency modes.
We find an excellent agreement between five distinct sets of results,
all showing that the use of evanescent pumping (gain) can compensate
the losses in the NRI slow-light regime.
\end{abstract}
\maketitle
The ability to stop and store optical pulses could usher in a range
of fundamentally new and revolutionary applications \cite{Milonni,Khurgin},
but the challenges encountered in our efforts to stop light are formidable.
This highly unusual state for an electromagnetic wave refers to the
situation where the lightwave completely stops despite the absence
of any dielectric or other barriers in the directions where it could
propagate. Unlike the confinement or trapping of light inside a dielectric
cavity, where one cannot spatially separate optical bits of information,
a stopped-light structure would allow for sequentially stopping and
storing spatially separated optical bits, thereby potentially leading
to all-optical memories \cite{Milonni,Khurgin}. Unfortunately, stopping
of light cannot be obtained with (non-switchable) periodic structures,
such as photonic crystals or coupled-resonator optical waveguides,
owing to their extreme sensitivity to disorder, which invariably destroys
the zero-group-velocity (ZGV) point(s) \cite{Khurgin}. With atomic
electromagnetically induced transparency one coherently imprints the
shape of an optical pulse into an electronic spin excitation, i.e.
light is {}``stored'' but not stopped because at the ZGV point all
photons are converted into atomic spins and light is completely extinguished
\cite{Lukin01}.

A method that could allow for true stopping of light in solid-state
structures and at ambient conditions was first suggested in 2007 \cite{Tsak07}.
Instead of relying on periodic back-reflections or on resonances,
the deceleration of light in this method is based on the use of negative
effective electromagnetic parameters in metamaterial waveguides, which
cause negative Goos-Hänchen phase-shift steps (i.e., deceleration)
in the propagation of a light ray. The negativity in the real parts
of the bulk constitutive parameters in metamaterials originates from
the response of deep-subwavelength elements or layers \cite{Hoffman07},
and can therefore be insensitive to disorder \cite{Singh10}. Hence,
it is anticipated that in metamaterial waveguides light can be coherently
decelerated and stopped even in the presence of disorder and surface
roughness. Indeed, time-domain simulations \cite{Tsak10,*Bai10,*Gan08}
reveal that stopping of light inside such, so called, {}``trapped
rainbow'' structures, is achievable despite deviations from the perfect
geometry (numerical {}``roughness'' at the interfaces) \cite{Zhao07}.
Moreover, a recent experimental work has demonstrated trapped rainbow
stopping of light in a tapered plasmonic system in the quasi-static
regime \cite{Smolyaninova10}.

A central task in this method of stopping and storing light is to
study whether the losses associated with the use of negative refractive
index (NRI) metamaterials can be overcome. Although a series of recent
works \cite{Xiao10,*Wuestner10,*Fang10} have shown that losses in
active \textquotedblleft{}fishnet\textquotedblright{} metamaterials
can be compensated, the considered structures were very thin in the
longitudinal direction, essentially being two-dimensional and spatially
dispersive. Furthermore, the dispersion relations and restrictions
obeyed by light in fishnet metamaterials are completely different
from those in trapped rainbow NRI waveguides. As a result, it is not
clear until now whether in such waveguides losses can, even in principle,
be overcome in a slow-light regime where the effective refractive
index experienced by light is negative \textendash{} even when use
is made of gain media \cite{Reza08,*Tsak08A}.

\begin{figure}
\includegraphics[width=0.45\textwidth]{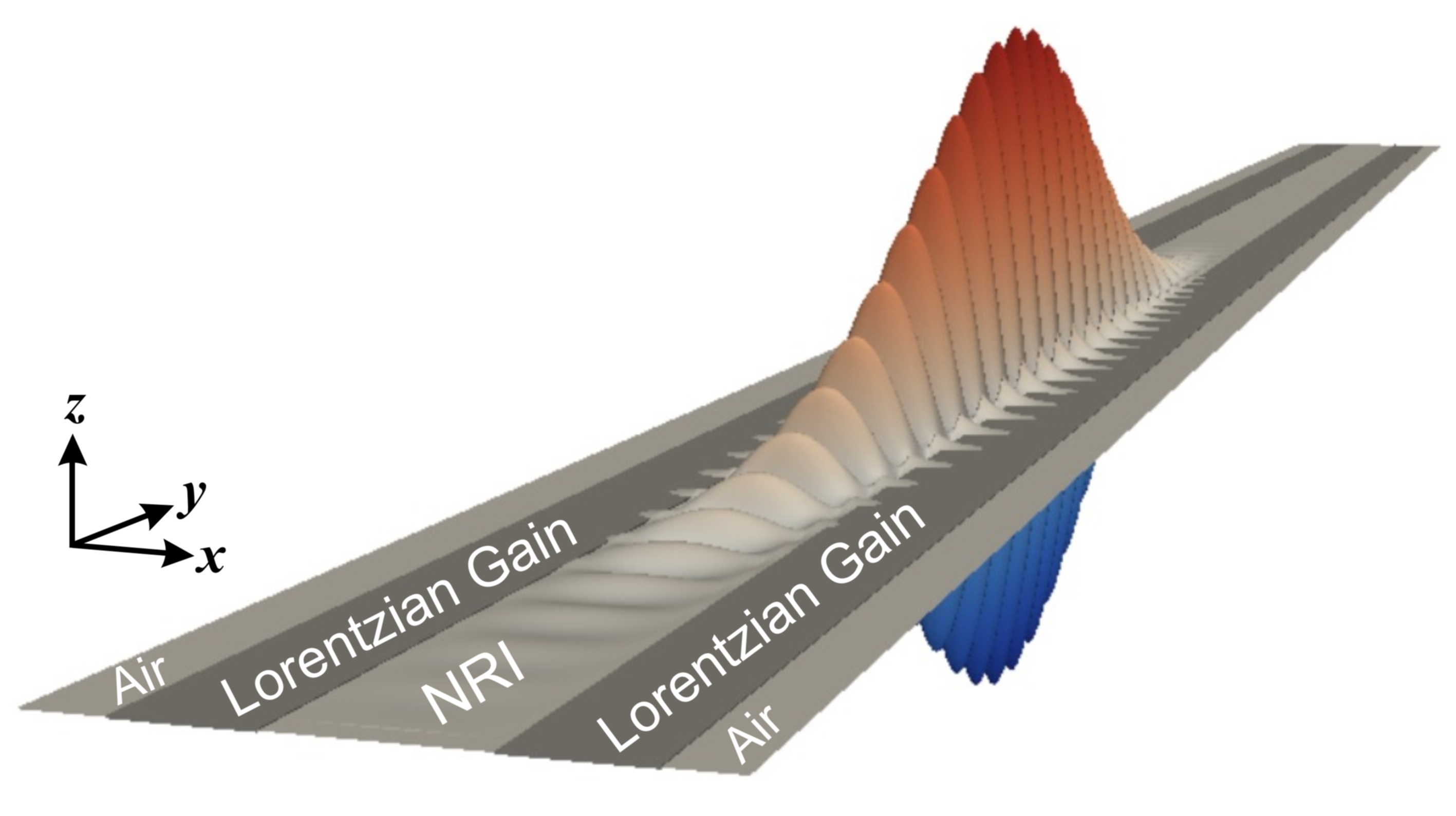}

\caption{\label{fig:Figure1} Illustration of the NRI slow-light heterostructure
considered in the analyses. Also shown is a characteristic snapshot
(from the FDTD simulations) of the propagation of the excited (dominant)
slow-light mode. See also \cite{Paper2-supplimentary2}.}

\end{figure}

In this Letter, on the basis of analytical calculations and rigorous
numerical simulations, we show how the incorporation of thin layers
of a gain medium in a passive trapped rainbow heterostructure can
compensate the losses while simultaneously preserving the negative
effective refractive index of the guided slow light. The specific
structure considered is illustrated in Fig.~\ref{fig:Figure1}. It
consists of a NRI core layer bounded symmetrically by two thin gain
layers that evanescently feed the supported guided modes \cite{Fang06}.
The whole structure is embedded in air. 

When gain or losses are present in one or more of the layers of a
heterostructure the supported guided modes become complex. Apart from
the usual complex-wavevector (real-frequency) solutions to the characteristic
equation, one may then also retrieve complex-frequency (real-wavevector)
solutions \cite{Alexander74,*Gammon74,*Kovener76,*Archambault09}.
For instance, it is well-known from prism-coupling to uniform metallic
films that fixing the incident light frequency and sweeping the angle
of incidence results in the excitation of complex-\emph{k} surface
plasmon polaritons (SPPs) exhibiting back-bending in the \emph{\textgreek{w}}-\emph{k}
dispersion diagram. By contrast, keeping the incident angle constant
and varying the frequency of the incident light gives rise to the
excitation of complex-\emph{\textgreek{w}} SPPs with distinct reflectivity
dips and no back-bending \cite{Alexander74,Kovener76}. The imaginary
part of a complex-\emph{\textgreek{w}} solution relates to the temporal
losses experienced by a light pulse \cite{Archambault09}. 

A numerical framework that is well-suited for the study of these modes
is the finite-difference time-domain (FDTD) method \cite{Taflove}.
With this method one can accurately launch the desired negative phase
velocity (backward) mode into the waveguide and directly investigate
its temporal losses \cite{Kirby09}. In our simulations we deploy
a modified total-field/scattered-field formulation \cite{Taflove},
with the excitation plane oriented perpendicularly to the central
axis of the heterostructure of Fig.~\ref{fig:Figure1}, and the amplitudes
of the $H_{z}$- and $E_{y}$- field components along the plane being
set to match the transverse profile of the backward TM$_{2}^{b}$
mode. The central frequency of the injected pulse is fixed to 400
THz ($\lambda_{0}$ = 750 nm), the side of the square FDTD cell has
a length of $\Delta x$ = $\lambda_{0}/200$ = 3.75 nm and the Courant
value is set to 0.7. The width of the core layer is $w_{c}$ = 0.35$\lambda_{0}$
= 262.5 nm, while the width of the gain layers in the cladding is
$w_{g}=0.25\lambda_{0}=187.5\text{ nm}$ whenever they are incorporated
in the heterostructure. We model the passive NRI of the core layer
using a broadband Drude response \cite[Chap. 1.3]{Engheta}: $n_{D}(\omega)=1-\omega_{p}^{2}/(\omega^{2}+i\omega\Gamma_{D})$,
with $\omega_{p}=2\pi\times893.8\times10^{12}$ rad/s and $\Gamma_{D}=0.27\times10^{12}$
rad/s. The frequency response of the permittivity of the gain layer
obeys a Lorentzian dispersion: $\epsilon_{L}(\omega)=\epsilon_{\infty}+\Delta\epsilon\omega_{L}^{2}/(\omega_{L}^{2}-i2\Gamma_{L}\omega-\omega^{2})$,
with $\epsilon_{\infty}$ = 1.001, $\Delta\epsilon$ = \textendash{}
0.0053, $\omega_{L}$ = $2\pi\times370\times10^{12}$ rad/s, and $\Gamma_{L}$
= $10^{14}$ rad/s, resulting in a line-shape that is similar to that
produced by, e.g., an electronic transition in a quantum dot \cite{Govyadinov07}. 

When the losses and gain are relatively small, as those used in the
considered NRI heterostructure, the imaginary part of the complex-\emph{\textgreek{w}}
solution is proportional to the imaginary part of the complex-\emph{k}
solution by a factor that is close-to-equal to the group velocity
\cite{Huang04}. Hence, in this case, one has an additional opportunity
(see Fig.~\ref{fig:Figure5} later on) to check the accuracy of the
obtained numerical results by examing whether such a proportionality
is fulfilled. Following the standard theory of active optical waveguides
\cite{Casey}, we assume that the saturation intensity for the gain
medium is sufficiently large, leading to a correspondingly large value
of the critical gain-length product beyond which gain depletion owing
to amplified spontaneous emission (ASE) can become significant \cite{MilonniLasers}.
Operation sufficiently below this limit implies that we are in the
linear regime where no gain depletion occurs, and that the effect
of ASE on the signal gain may be disregarded, in accordance with analogous
studies of active optical structures \cite{Casey,Jiang09,*Lu10,Zheludev08}.

To analyically confirm the accuracy of the numerical results we have
developed a frequency-domain transfer-matrix method (TMM) capable
of identifying both the complex-\emph{\textgreek{w}} and complex-\emph{k}
modes of the multilayer NRI heterostructure. The method derives the
dispersion equation and uses the argument principle method to locate
and isolate its zeros either on the complex-\emph{\textgreek{w}} or
the complex-\emph{k} plane \cite{Kwon04}. Suitable conformal mappings
are deployed in both cases to unfold the four-sheeted Riemann surfaces
associated with the characteristic equations. Upon isolation of a
zero, the Newton-Raphson method is used to pinpoint and track its
location on the complex plane. The technical details of the overall
procedure will be presented elsewhere. 

In our analyses we consider the following four cases: (I) neither
loss in the NRI core layer ($\Gamma_{D}$ = 0) nor gain in the cladding
layers (the cladding is only air), (II) loss in the NRI core layer
but no gain in the cladding, (III) both loss in the NRI region and
gain in the cladding strips, and (IV) the NRI core layer is modeled
as being lossless and gain is used in the two cladding layers.

\begin{figure}
\includegraphics[width=0.49\textwidth]{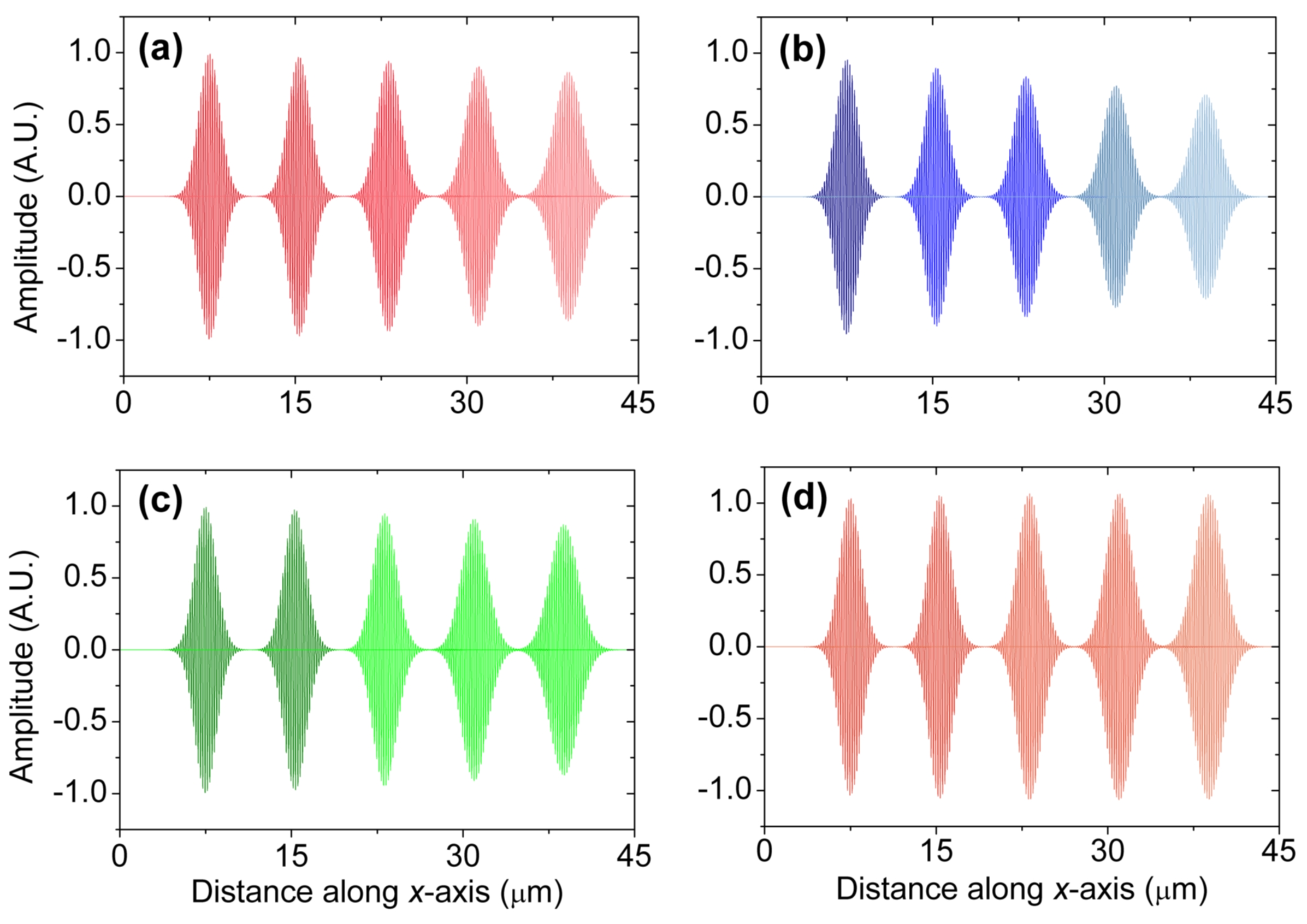}

\caption{\label{fig:Figure4-1} Snapshots of slow-light pulse propagation along
the central axis of the considered waveguides for: (a) case I (neither
loss nor gain); (b) case II (loss but no gain); (c) case III (both,
loss and gain); and (d) case IV (gain but no loss). The lighter the
color, the later in time the snapshot is taken.}

\end{figure}

In order to validate the causal dynamics of the injected pulse with
the FDTD method we created an animation of its propagation along the
considered heterostructure \cite{Paper2-supplimentary2}. From there
one may directly see that first, a single slow-light guided mode is
excited and second, the mode experiences an effective refractive index
$n_{\textit{eff}}$ with a negative real part, having antiparallel
phase and group velocities.

\begin{figure}
\includegraphics[width=0.45\textwidth]{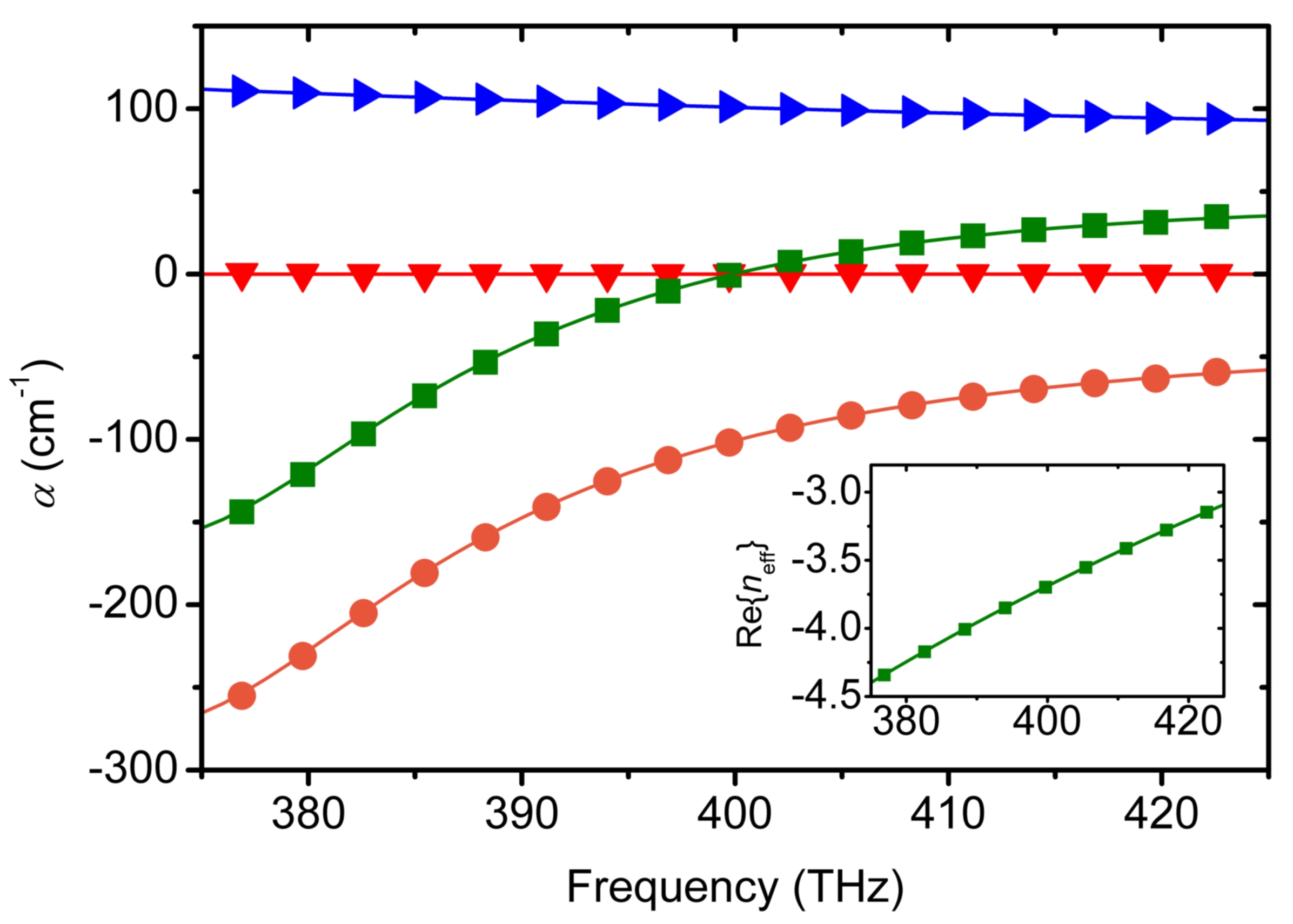}

\caption{\label{fig:Figure3-1} Comparison between FDTD (symbols) and TMM (lines)
calculations of the absorption coefficient $\alpha$ (spatial losses)
versus frequency for the TM$_{2}^{b}$ mode in case I (red), case
II (blue), case III (green) and case IV (orange). The inset depicts
the frequency dispersion of Re$\{\textit{n}_{\textit{eff}}\}$ in
all four cases. }

\end{figure}

\begin{figure}
\includegraphics[width=0.49\textwidth]{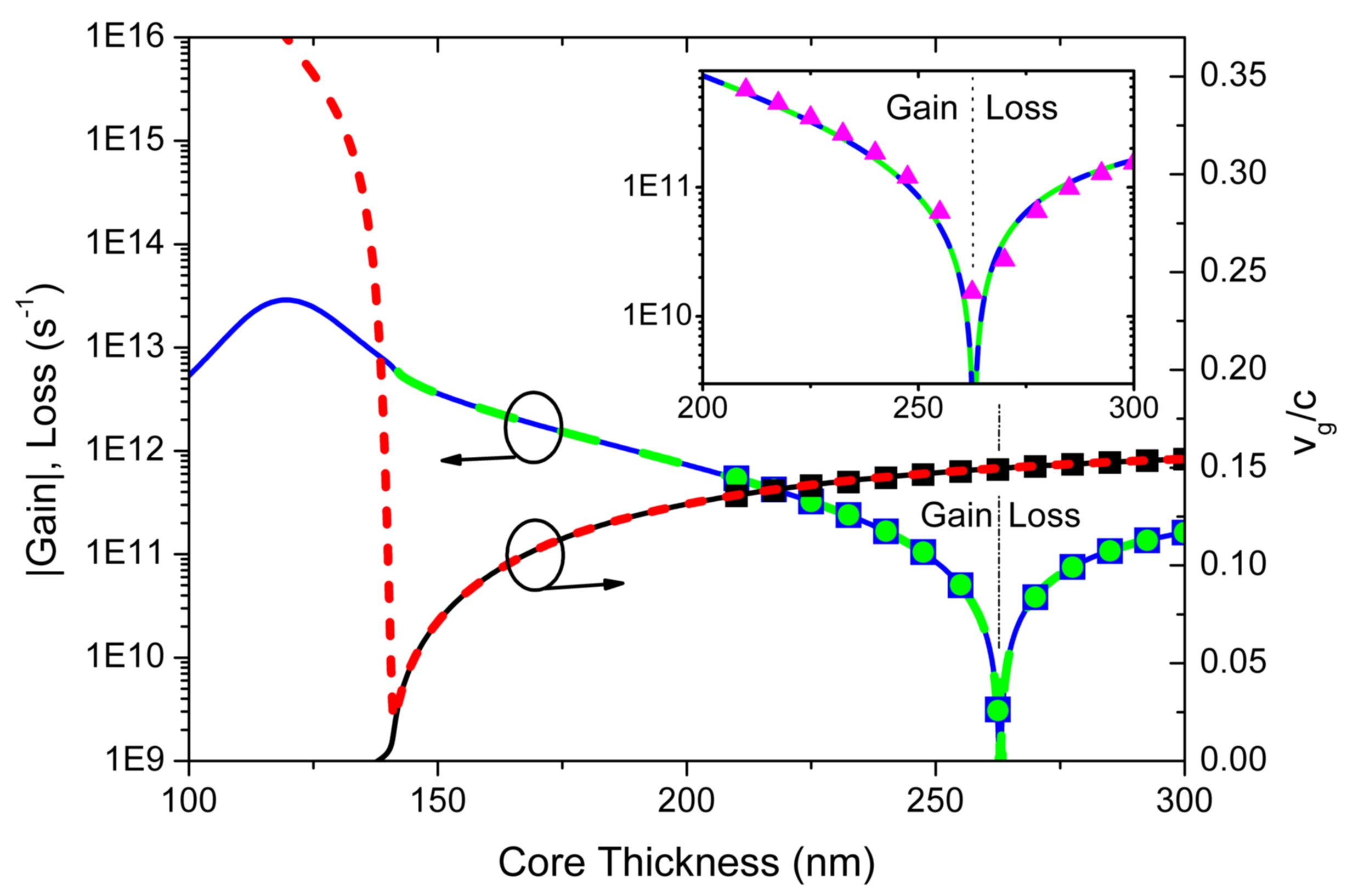}

\caption{\label{fig:Figure5}(color online) Comparison between FDTD (symbols)
and TMM (lines) calculations of the temporal losses/gain and group
velocity of the complex-\emph{\textgreek{w}} and complex-\emph{k}
solutions with varying core thickness (case III). Shown are the group
velocity ($v_{g}$) of the complex-\emph{\textgreek{w}} solutions
(black), the group velocity of the complex-\emph{k} solutions (red
dashed line), the imaginary part of the complex-\emph{\textgreek{w}}
solutions (blue) and the imaginary part of the complex-\emph{k} solutions
multiplied by \emph{$v_{g}$} (green). The inset shows the rate of
energy loss (or gain) for the whole wavepacket (purple symbols) with
varying core thickness as calculated by the discrete Poynting's theorem
within the FDTD method.}

\end{figure}

Sucessive snapshots of the Gaussian pulse propagating down the NRI
waveguide for cases I-IV are depicted in Fig.~\ref{fig:Figure4-1}.
We see that for case I (neither loss nor gain) the amplitude of the
guided pulse decays with distance {[}Fig.~\ref{fig:Figure4-1}(a){]}.
This reduction is not a result of the pulse losing energy but arises
entirely owing to group-velocity dispersion, which causes the guided
slow pulse to broaden, thereby leading to a gradual decay in amplitude.
Figure~\ref{fig:Figure3-1} (red solid line and symbols) confirms
the fact that energy is conserved, as the absorption coefficient $\alpha=2\omega\text{Im}\{\textit{n}_{\textit{eff}}\}/c$
(spatial losses) is zero throughout the frequency spectrum of the
pulse. The complex $n_{\textit{eff}}$ shown in Fig.~\ref{fig:Figure3-1}
for all considered cases is extracted by recording the amplitude of
the pulse at two fixed points along the central axis of the waveguide
over time, and then dividing the Fourier transforms of the two time
series \cite{Taflove,Kirby09}. When dissipative loss is introduced
into the NRI core (case II) the amplitude of the pulse decreases even
further compared to case I {[}Fig.~\ref{fig:Figure4-1}(b){]}. Figure~\ref{fig:Figure3-1}
(blue solid line and symbols) shows that now Im$\{\textit{n}_{\textit{eff}}\}>0$
for all frequencies, as expected.

By introducing gain in the cladding layers \textit{\emph{(case III)
we evanescently pump the pulse }}\cite{Fang06} and allow for the
compensation of its propagation losses {[}compare Fig.~\ref{fig:Figure4-1}(c)
with Fig.~\ref{fig:Figure4-1}(a){]}. Indeed, Fig.~\ref{fig:Figure3-1}
(green solid line and squares) shows that at approximately 400 THz
(central frequency of the pulse) the imaginary part of the effective
refractive index becomes zero, while for smaller frequencies Im$\{\textit{n}_{\textit{eff}}\}$
assumes \textit{\emph{negative}} values (amplification). Thus, in
case III there is a continuous range of frequencies ($f<400\text{ THz}$)
where we simultaneously have Re$\{\textit{n}_{\textit{eff}}\}<0$
(inset in Fig.~\ref{fig:Figure3-1}) and Im$\{\textit{n}_{\textit{eff}}\}<0$
(green line for $f<400$ THz in Fig.~\ref{fig:Figure3-1}). We note
that the optogeometric parameters of the heterostructure have been
chosen such that a light-pulse experiences almost the same frequency
dispersion for all cases, with differences between the various Re$\{\textit{n}_{\textit{eff}}\}$
(cases I-IV) being indiscernible at the linear scale of the inset
in Fig.~\ref{fig:Figure3-1}. For all cases presented we find that
the parameters retrieved from the FDTD simulations (symbols) are in
excellent agreement with those calculated using the TMM (lines).

To further confirm that light amplification is in principle possible
in the negative index slow-light regime we {}``switch off'' the
losses, while maintaining the gain in the cladding strips. Figure~\ref{fig:Figure4-1}(d)
shows that in this case (IV) the negative-phase-velocity slow pulse
is amplified while propagating along the waveguide. In particular,
it is seen that the pulse amplitude at around \emph{x} = 40 \textgreek{m}m
exceeds its initial amplitude despite the fact that the pulse has
been broadened due to group-velocity dispersion. This conclusion is
further confirmed by finding that Im$\{\textit{n}_{\textit{eff}}\}<0$
throughout the spectrum of the Gaussian pulse, as shown in Fig.~\ref{fig:Figure3-1}
(orange solid line and symbols).

Next, we examine how the spatial and temporal losses (or gain) experienced
by both the central frequency of the pulse and the pulse as a whole
vary with core thickness (Fig.~\ref{fig:Figure5}). The complex-\emph{\textgreek{w}}
solutions can be calculated with the FDTD method by recording the
spatial variation of the field amplitude along the central axis of
the heterostructure at two different time points, and then dividing
the spatial Fourier transforms of the two longitudinal spatial profiles.
The rate of energy change for the whole wavepacket (total loss or
gain) is calculated using the discrete Poynting's theorem integrated
over a spatial region sufficiently wide to contain the pulse. 

Figure~\ref{fig:Figure5} shows that for core thicknesses above 262
nm the central frequency of the pulse experiences loss. For smaller
thicknesses, for which the amplitude of the field increases inside
the gain region, we find that the gain supplied by the cladding strips
overcompensates the loss induced by the core layer. At a core thickness
of 262 nm the central frequency experiences neither gain nor loss,
while the wavepacket as a whole experiences gain (inset in Fig.~\ref{fig:Figure5}).
In all cases we have verified that Re$\{\textit{n}_{\textit{eff}}\}<0$
(data not shown here). 

Overall, we find excellent agreement and consistency between five
distinct sets of results: the spatial losses/gain (multiplied by the
group velocity \cite{Huang04}) for the central frequency as calculated
by the FDTD (green dots) and the TMM (green dashed line), the temporal
losses/gain for the central frequency as calculated by the FDTD (blue
squares) and TMM (blue dashed line), and the temporal losses of the
whole wavepacket as calculated by the FDTD method (purple symbols
in the inset of Fig.~\ref{fig:Figure5}). This fact provides further
evidence that loss compensation is in principle possible in the slow-light
NRI regime. 

Finally, we note that for core thicknesses smaller than around 140
nm the group velocity of the complex-\emph{k} mode characteristically
differs from that of the complex\emph{-\textgreek{w}} mode (red dashed
and black solid lines in Fig.~\ref{fig:Figure5}). As with the case
of SPPs in plasmonic films \cite{Alexander74,Kovener76}, the group
velocity of the complex-\emph{k} solutions exhibits a \textquotedblleft{}back-bending\textquotedblright{},
never becoming zero, while that associated with the complex\emph{-\textgreek{w}}
solutions may reduce to zero even in the presence of excessive gain
(or losses).

In conclusion, by studying in the time- and frequency-domain the complex-\emph{\textgreek{w}}
and complex-\emph{k} modes in NRI slow-light waveguides with gain
in the cladding region, we have shown that it is possible to compensate
the dissipative optical losses. This geometry allows for lossless
or amplified slow-light propagation in the regime where the real part
of the effective refractive index experienced by the guided modes
remains negative. We believe that this work could aid the realization
of lossless metamaterial waveguides to be used in a wealth of photonic
and quantum optics applications.

We gratefully acknowledge financial support provided by the EPSRC
and the Royal Academy of Engineering.

\newpage{}

\bibliographystyle{apsrev4-1}
\bibliography{Transfer-bib}

\end{document}